# Electrical switching and oscillations in vanadium dioxide

Alexander Pergament[*], Andrey Velichko, Maksim Belyaev, Vadim Putrolaynen

*Institute of Physics and Technology, Petrozavodsk State University, 185910 Petrozavodsk, Russia*

[*]Corresponding author: aperg@psu.karelia.ru, Petrozavodsk State University, Petrozavodsk, 185910, Russia.

**Abstract.** We have studied electrical switching with S-shaped *I-V* characteristics in two-terminal MOM devices based on vanadium dioxide thin films. The switching effect is associated with the metal-insulator phase transition. Relaxation oscillations are observed in circuits with $VO_2$-based switches. Dependences of the oscillator critical frequency $F_{max}$, threshold power and voltage, as well as the time of current rise, on the switching structure size are obtained by numerical simulation. The empirical dependence of the threshold voltage on the switching region dimensions and film thickness is found. It is shown that, for the $VO_2$ channel sizes of 10×10 nm, $F_{max}$ can reach the value of 300 MHz at a film thickness of ~20 nm. Next, it is shown that oscillatory neural networks can be implemented on the basis of coupled $VO_2$ oscillators. For the weak capacitive coupling, we revealed the dependence of the phase difference upon synchronization on the coupling capacitance value. When the switches are scaled down, the limiting time of synchronization is reduced to $T_s$ ~13 μs, and the number of oscillation periods for the entering to the synchronization mode remains constant, $N_s$ ~ 17. In the case of weak thermal coupling in the synchronization mode, we observe in-phase behavior of oscillators, and there is a certain range of parameters of the supply current, in which the synchronization effect becomes possible. With a decrease in dimensions, a decrease in the thermal coupling action radius is observed, which can vary in the range from 0.5 to 50 μm for structures with characteristic dimensions of 0.1 to 5 μm, respectively. Thermal coupling may have a promising effect for realization of a 3D integrated oscillatory neural network.

*Keywords*: metal-insulator transition, vanadium dioxide, electronic switching, electrical oscillations, artificial neural network.

**1. Introduction**

Vanadium dioxide is an archetypal strongly correlated system that undergoes a Mott metal-insulator transition (MIT) due to the electron interactions accompanied by a structural Peierls transition [1-3]. The switching effect in vanadium dioxide is associated with the MIT, and oscillations are observed in circuits with metal/VO$_2$/metal switching structures [1]. In the present paper, we report on the numerical simulation and experimental study of the electrical switching effect when scaling the channel region, as well as on the switching dynamics of coupled VO$_2$-based oscillators with a capacitive and thermal coupling, and explore the capability of their application in an oscillatory neural network (ONN).

An artificial neural network is one of the most promising approaches for development of the next generation computing architectures realizing the brain-inspired massively parallel computing paradigm [3]. Hardware implementations of ONNs are based both on current CMOS devices (e.g., phase-locked loop circuits [4] or Van der Pol oscillators [5]) and on emerging new devices, such as, spin-torque nano-oscillators [6], switches based on materials with metal-to-insulator transitions [1, 3, 7] or charge-density-wave transitions [8], and oxide RRAM [9].

Vanadium dioxide is currently considered as one of the key materials for neuromorphic oxide electronics [3]. Two-terminal thin-film metal/oxide/metal devices based on VO$_2$ exhibit S-type switching, and in an electrical circuit containing such a switching device, relaxation oscillations are observed under certain conditions, namely, when the load line intersects the *I-V* curve at a unique point in the negative differential resistance (NDR) region [1]. The switching effect in vanadium dioxide is caused by the MIT occurring in this material at $T_t$ = 340 K [2, 10-12]. This thermally driven MIT between the insulating monoclinic and the metallic rutile phases of VO$_2$ most likely belongs to the class of electronic Mott transitions [10], since the energy gap in the insulating phase arises due to Mott-Hubbard type Coulomb correlations [11]. Nevertheless, the switching effect in relatively low electric fields (namely, if the threshold switching field does not reach a value at which the development of high-field effects becomes feasible [12, 13]) is purely thermal. Basically, switching is governed by the current-induced Joule heating, up to the transition temperature $T_t$, of the oxide film between two metal electrodes.

In an ONN, an elementary cell comprises an oscillator circuit, and the cells are locally coupled by resistors or capacitors [7-9]. Also, in the work [5], the variable resistive coupling via a memristive device has been proposed. Factually, such a type of coupling might be controllable, since the transition of the memristor between the ON and OFF states would bring about the coupling strength hopping between the strong-coupling and weak-coupling modes. Note, that the $VO_2$-based oscillators can be linked to each other by a thermal coupling owing to a heat-induced switching mechanism. In this case, the coupling strength could also be made controllable via the variation of the heating intensity.

Capacitive and thermal couplings allow obtaining DC decoupling of two oscillators. DC decoupling does not shift operating points of oscillator schemes and is a basic requirement for oscillators' stable operation under variation of coupling forces. Nevertheless, the question which one of these couplings is more promising for ONN realization and how switch scaling affects the oscillator's main parameters remains unclear.

Thus, the objective of this work is to study the switching dynamics as a function of oscillator size and explore the capability of their application in ONNs.

**2. Experimental techniques and numerical calculation methods**

Vanadium dioxide films were deposited by DC magnetron sputtering onto *R*-cut sapphire substrates using an AJA Orion 5 sputtering system. The film deposition process has been described in more detail elsewhere [7]. The oxide film thickness was measured by a NTEGRA Prima atomic force microscope (AFM), and it was typically in the range of 200 to 250 nm.

Planar devices were formed by the optical lithography (a Heidelberg Instruments µPG 101 laser lithograph) and lift-off processes. The electrodes were two-layer V-Au metal contacts with the overall thickness of about 50 nm (30 nm of vanadium and 20 nm of gold). After lithography, the structures were annealed in air at a temperature of 380 °C for 10 min. To study the effect of thermal coupling, electrically isolated planar microstructures were formed at a distance of $d = 21$ µm from each other, with the shape of the contacts shown in Fig. 1a. The length $l$ and gap $h$ of the switch interelectrode space were ~ 3-4 µm and 2.5 µm, respectively.

The X-ray structural analysis showed that annealing was accompanied by partial oxidation and crystallization of the vanadium oxide films with formation of $V_2O_5$, $V_2O_3$ and, predominantly, $VO_2$ phases. This was also confirmed by the temperature dependence of the film resistivity (Fig. 2b) measured by the four-probe method, which demonstrated a resistivity jump of ~$10^2$ at the transition temperature, $T_t$ ~320 K. The temperature coefficient of resistance (TCR) at room temperature, calculated from the graph of Fig. 1b, was ~2.1 %/K.

Note, that the obtained value of $T_t$ is lower than 340 K, which is typical for thin films as compared to perfect single crystals [13, 14].

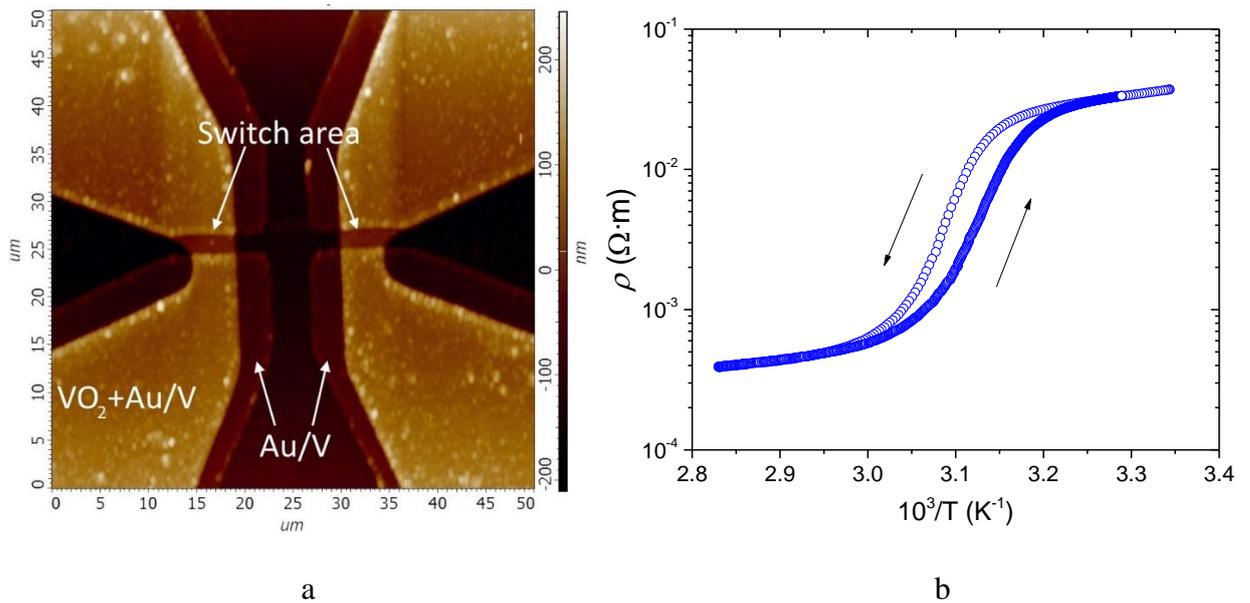

a

b

**Fig. 1.** (a) AFM image of the structure under study and (b) the temperature dependence of the film resistivity.

The study of oscillation dynamics was performed with a four-channel oscilloscope Picoscope 5442B with the maximum sampling rate of 125 MS/sec in 14-bit mode. A two-channel sourcemeter Keythley 2636A was used for the DC *I-V* characteristic measurements (sweeping rate 1V/s), and also as an oscillatory circuit voltage and current source.

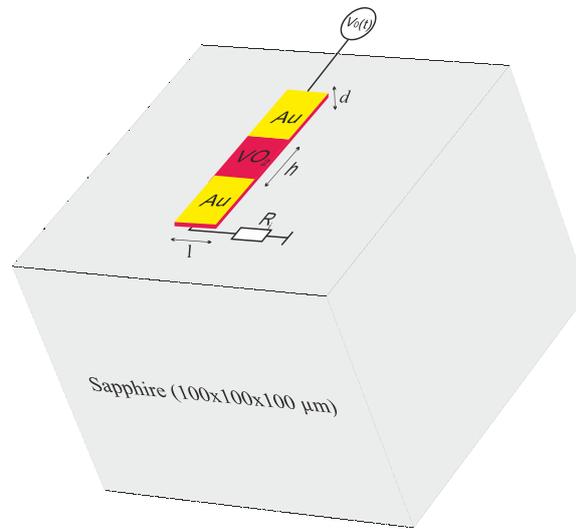

**Fig. 2.** Model structure image with the parameters $h = l = 1$ µm, and $d = 100$ nm.

The simulation was performed using the COMSOL Multiphysics software platform. The calculation area consisted of a 100×100×100 µm sapphire substrate element with a switch on the lateral side, which in turn consisted of a $VO_2$ film strip and gold film electrodes, while the other faces were maintained at room temperature $T_0 = 300$ K. The structure image with the indicated switch width *l* and interelectrode gap *h*, as well as the size of the gold contacts, is shown in Fig. 2. In the case when the dimensions were the same, we applied the definition $h = l = a$. The dimensions of contacts *a* and film thicknesses *d* are given in Table 1. A sawtooth voltage with a forward and backward duration with sweeping rate 1V/s was applied to the Au contacts, and the amplitude was selected experimentally, depending on the switch threshold voltage. The electrical and thermal parameters of the materials for modeling were taken from the works [15, 16]. The resistivity versus temperature curve corresponded to the direct (when heated) branch of the dependence shown in Fig. 1b, and the function was extrapolated by a constant value beyond the experimental values. Simulation of the operation of electrical circuits was performed in the LTSpice software. A current-limiting resistor $R_i = 250$ Ω was connected in series to the switch in the electrical circuit in experiments with oscillator circuits.

## 3. Results and discussion

### 3.1. Simulation of the switch *I-V* characteristic in nanoscale

One of the important points in the study and design of the switching element based on $VO_2$ is the modeling of its physical properties and the estimation of critical parameters depending on the configuration and size of the channel area. The key parameters include the threshold voltages $V_{th}$ (switching-on) and $V_h$ (holding), the corresponding threshold currents ($I_{th}$, $I_h$), the maximum temperature reached at the moment of switching $T_{max}$, the time of the transition from the high-resistance state to the low-resistance state $\tau_{on}$, and the reverse transition time $\tau_{off}$, as well as the switching-on power consumption $P_{th}$.

A typical DC current-voltage characteristic of a planar switch (Fig. 1a), measured in the sweep voltage mode, is shown in Fig. 3 (experimental curve). A sharp increase in current of ~25 times and the transition to a low-resistance state are observed at threshold values $V_{th} \sim 5.6$ V and $I_{th} = 0.51$ mA, and the reverse transition at $V_h \sim 2.2$ V and $I_h = 1.26$ mA. The dynamic resistances of the ON and OFF states are $R_{on} \sim 350$ Ω and $R_{off} \sim 12$ kΩ, respectively.

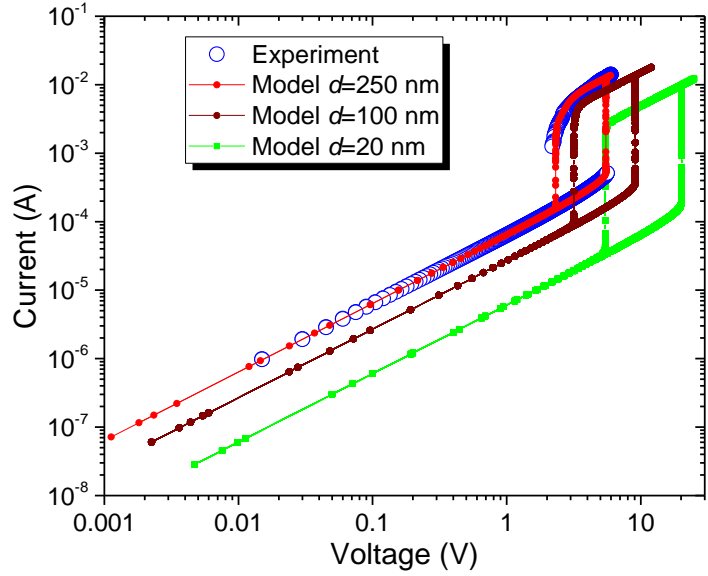

**Fig. 3.** DC *I-V* characteristic of $VO_2$-based switch ($R_i$=250 Ω): experimental curve and model ones at $d$ = 250, 100, and 20 nm.

The simulation results, taking into account the actual dimensions of the switch ($l$ = 3-4 μm and $h$ = 2.5 μm), for $d$ = 250 nm are shown in Fig. 3 (model curves). One can see that they describe the experimental *I-V* characteristic with a 90% accuracy, which indicates the sufficiency of the numerical model based on the classical heat-transfer and heat-release laws used in the Comsol simulation. For comparison, the *I-V* characteristics for a structure with thicknesses of $VO_2$ films $d$ = 100 nm and 20 nm at $R_i$ = 250 Ω have been calculated in a similar way. We note that the threshold voltages ($V_{th}$, $V_h$), as well as the resistances $R_{on}$ and $R_{off}$, increase with decreasing $d$. In order to derive a formula-expressed dependence, we have performed a number of model calculations with unified dimensions, namely at $h = l = a$.

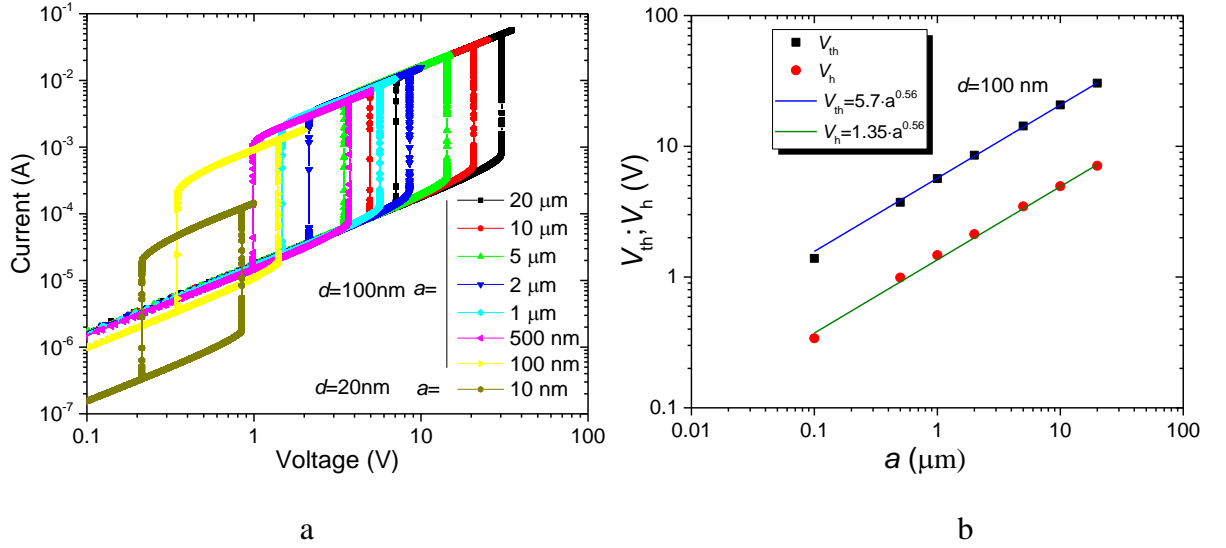

a

b

**Fig. 4.** a) Calculated *I-V* curves depending on the structure dimensions ($a$ = 0.1-20 μm) at a constant film thickness value $d$ = 100 nm (in case of $a$ = 0.01 μm, $d$ = 20 nm). b) Threshold voltages as functions of $a$ and their numerical approximation.

Figure 4a shows a set of model *I-V* curves for a switching region size $a$ variation in the range of 0.1 to 20 μm at a constant film thickness $d$ = 100 nm ($R_i$ = 0 Ω), and, for comparison, an *I-V* characteristic of a nanosized structure with $a$ = 10 nm and $d$ = 20 nm. The logarithmic scale of the ordinate axes makes it possible to see the regularity and the wide range of variation, by almost two orders of magnitude, of the threshold currents and voltages observed when scaling. The dependences $V_{th}(a)$ and $V_h(a)$, shown in Figure 4b in bilogarithmic coordinates, are linear and parallel to each other. Based on this observation, one can propose an empirical approximation for the threshold voltages:

$$V_{th} = \frac{57}{\sqrt{d}} \cdot a^{0.56} , \qquad (1)$$

$$V_h = \frac{13.5}{\sqrt{d}} \cdot a^{0.56} , \qquad (2)$$

where $d$ is the film thickness (nm), and $a$ is the switching channel size (μm).

Physical interpretation of (1) – (2) formulae becomes clear if they are derived analytically from the simplest assumptions. If we assume that at the moment of switching from OFF to ON state at $V=V_{th}$, and in the equilibrium condition the Joule energy released on $VO_2$ channel is completely transmitted to the substrate owning to heat-transfer process (Fourier law) then we can write:

$$\frac{V_{th} \cdot V_{th}}{R} = \lambda \cdot S \cdot (T_{th} - T_0), \tag{3}$$

where $R$ - is the channel resistance, $\lambda$ - is the heat-transfer coefficient, $(T_{th}-T_0)$ - maximum temperature difference, $S$ - is the channel's effective square that can be expressed as $S=(l \cdot h)^\beta$, where $\beta$ - is the exponential coefficient that determines the effective area of the heated zone in the interelectrode gap (it is evident that $\beta<1$). If $l=h=a$, the resistance can be expressed as $R=\rho_{off} \cdot l/(h \cdot d)=\rho_{off}/d$, where $\rho_{off}$ - is the channel's effective resistivity at the high impedance state at the switching moment and the square is $S=a^{2 \cdot \beta}$ If to take these into account we can reconstruct formula (3) by expressing $V_{th}$ as

$$V_{th} = \frac{\sqrt{\lambda \cdot \rho_{off} \cdot (T_{th} - T_0)}}{\sqrt{d}} \cdot a^\beta. \tag{4}$$

Therefore, by simplest assumptions we obtain formula (4) that is analogous to formula (1) and clarifies the physics of empirically found coefficients and exponential dependencies on $a$ and $d$. The similar analytical approximation could be also obtained for formula (2); it will differ only by resistivity at the low impedance state $\rho_{on}$ and is given by:

$$V_h = \frac{\sqrt{\lambda \cdot \rho_{on} \cdot (T_{th} - T_0)}}{\sqrt{d}} \cdot a^\beta \tag{5}$$

From this it follows that the threshold voltage ratio at unchanged $T_0$ is just the function of resistivity:

$$\frac{V_{th}}{V_h} = \sqrt{\frac{\rho_{off}}{\rho_{on}}} \qquad (6)$$

Formula (6) shows that the higher $\rho_{off}/\rho_{on}$ ration is the more different the threshold voltages are. According to empirical formulas (1)-(2) the ratio of threshold voltages is $V_{th}/V_h =$ const $\approx 4.22$, regardless of $d$ and $a$. Analytical formula (6) provides similar result $V_{th}/V_h \sim 5$, if we take the values for $\rho$ from temperature dependence (Fig 1.b) at the boundaries of hysteresis loop (at the inflection points).

Table 1 presents the threshold voltage estimates according to these formulas. Formulas (1) and (2) are of interest as they allow us to accurately estimate the threshold voltages over a wide range of scale changes provided that $a \geq d$ at a specified resistivity temperature dependence (see Fig. 1b). For $a < d$, the deviation from Equations (1) and (2) increases due to a more complex distribution of the current density over the film thickness, which differs from the homogeneous one. Apparently, in a nanometer (1-20 nm) channel size range, the physics of the switching effect is complicated by nanosized tunneling phenomena, and a simple thermal model would not be sufficient to describe the actual experimental data. Moreover, the MIT can be suppressed at $a < 2$ nm due to the dimensional effects [13], because under equilibrium conditions, the MIT degenerates at scales comparable to correlation length. For vanadium dioxide, this correlation length has been estimated to be $\xi \sim 2$ nm [15].

**Table 1**. Parameters of modeled *I-V* characteristic depending on the scaling dimension $a$ and film thickness $d$.

| $a$, μm | $d$, nm | $V_{th}$, V model | $V_{th}$, V (1) | $V_h$, V model | $V_h$, V (2) | $I_{th}$, A | $I_h$, A | $\tau_f$, ns | $\tau_{off}$, ns | $P_{th}$, mW |
|---|---|---|---|---|---|---|---|---|---|---|
| 20 | 100 | 30.41 | 30.51 | 7.1 | 7.22 | $6.9 \cdot 10^{-4}$ | 0.0094 | 152 | 14063 | 25.02 |
| 10 | 100 | 20.71 | 20.69 | 4.95 | 4.9 | $4.8 \cdot 10^{-4}$ | 0.0063 | 42 | 5266 | 11.48 |
| 5 | 100 | 14.3 | 14.03 | 3.47 | 3.32 | $3 \cdot 10^{-4}$ | 0.0041 | 19.6 | 2290 | 5.41 |
| 2 | 100 | 8.55 | 8.40 | 2.13 | 1.99 | $1.8 \cdot 10^{-4}$ | 0.0029 | 6.8 | 691 | 1.83 |
| 1 | 100 | 5.66 | 5.7 | 1.47 | 1.35 | $1.1 \cdot 10^{-4}$ | 0.0018 | 2.8 | 241 | 0.77 |

| | | | | | | | | | |
|---|---|---|---|---|---|---|---|---|---|
| 0.5 | 100 | 3.73 | 3.86 | 0.99 | 0.91 | $6.8 \cdot 10^{-5}$ | $8.2 \cdot 10^{-4}$ | 1.15 | 130 | 0.31 |
| 0.1 | 100 | 1.39 | 1.56 | 0.34 | 0.37 | $1.8 \cdot 10^{-5}$ | $1.5 \cdot 10^{-4}$ | 0.88 | 52 | 0.027 |
| 0.01 | 20 | 0.85 | 0.96 | 0.21 | 0.22 | $1.8 \cdot 10^{-6}$ | $2 \cdot 10^{-5}$ | 0.04 | 3,2 | 0.0015 |
| 1 | 20 | 14.17 | 12.74 | 3.48 | 3.01 | $7.6 \cdot 10^{-5}$ | $9.5 \cdot 10^{-4}$ | 0.75 | 179 | 1.07 |
| 10 | 20 | 45.36 | 46.27 | 10.92 | 10.96 | $2.7 \cdot 10^{-5}$ | 0.0031 | 29.9 | 5720 | 12.47 |

As has been shown in [7], the value of $\tau_{on}$ is determined by the sum of the delay time $\tau_{del}$ and the time of the switching development $\tau_f$ ($\tau_{on} = \tau_{del} + \tau_f$). The time $\tau_{del}$ can vary over a wide range depending on the loading conditions (for example, the signal amplitude or the voltage rise rate) and it is determined by the time of the structure heating up to a certain critical temperature corresponding to the commencement of the conductivity growth due to the MIT. The front time $\tau_f$ is the time for which the current channel is formed and a sharp increase in current is observed on the current-voltage characteristic. This value defines the minimum value of $\tau_{on}$ and it depends only on the size of the channel area. Therefore, we present the results just for the $\tau_f$ values in Table 1 obtained from modeled *I-V* characteristics. One can see that, as the scaling parameter *a* decreases, the switching development time significantly falls, and in logarithmic coordinates, it also demonstrates a linear dependence. The transition time from the low-resistance to the high-resistance state $\tau_{off}$ is associated with cooling of the switching channel region and it considerably exceeds $\tau_f$, all other conditions being equal. The value of $\tau_{off}$ also falls with a decrease in the *a* dimension (see Table 1). The threshold power $P_{th}$, released in the switching structure at the switching-on moment, varies with scaling and decreases with a decrease in *a* (Table 1, last column).

### 3.2. Study of single-oscillator circuit based on VO$_2$ switch in nanoscale

To observe auto-oscillations based on a VO$_2$ switch, two circuit types are usually used, namely, those with a current source $I_0$ and voltage source $V_0$ (see Fig. 5). The combination of $V_0$ and load resistance $R_L$ is the analogue to the current source $I_0$, therefore it does not matter which variant to prefer. The load resistance $R_L$ and power supply voltage $V_0$ or current $I_0$ are chosen so that the load line passes through the NDR region, i.e. the region of currents and voltages between the points of switching-on and off. This is a condition for the oscillations to occur, and the current resistor value $R_i$ is negligible as compared to the resistance of the whole circuit. The supply current $I_0$ value is usually chosen in the middle of the NDR region between $I_{th}$ and $I_h$, $I_0=I_{NDR}=(I_h+I_{th})/2$. Typical voltage and current oscillation waveforms obtained experimentally (for the scheme in Fig. 5.a) are shown in Fig. 6.

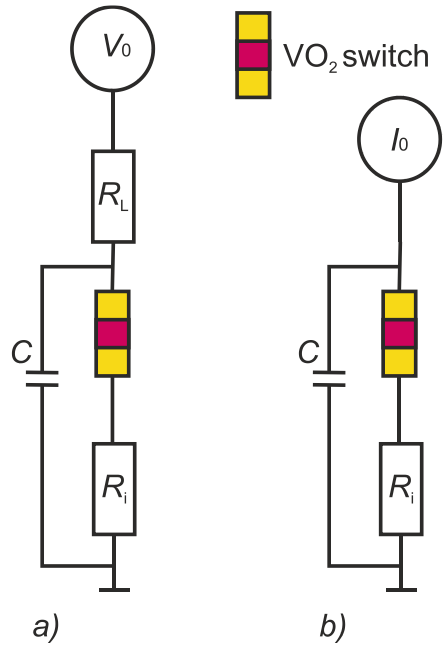

**Fig. 5**. Single oscillator circuits a) with a voltage source $V_0$ and b) with a current source $I_0$.

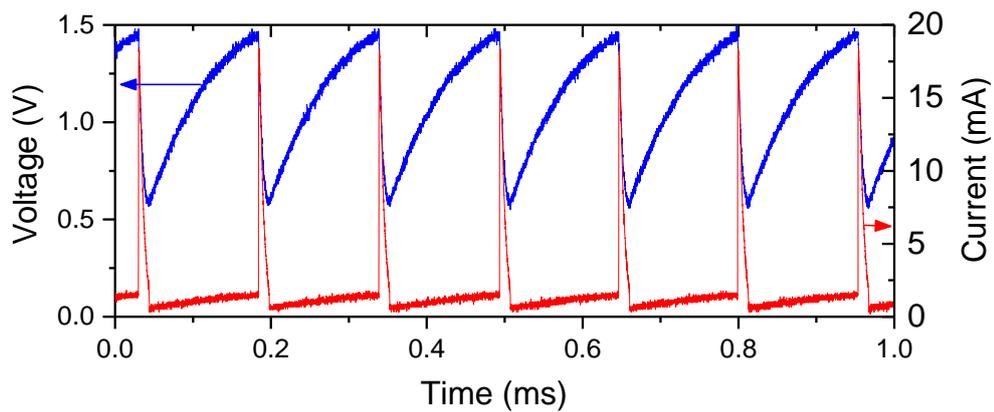

**Fig. 6**. Experimental relaxation oscillations of voltage (blue) and current (red) for a single oscillator, for the scheme in Fig.5.a ($R_L$=50 k, $V_0$=82 V, $R_i$ =10 Ω, $C$=100 nF, $V_{th}$=1.4 V, $V_h$=0.6 V, $R_{off}$=1.1 kΩ, $R_{on}$=40 Ω).

For the circuit in Fig. 5a, the natural oscillation frequency can be evaluated from the equation [1, 17]:

$$F_0 \sim \left[ \frac{CR_L}{x} \ln\left[ \frac{V_0 - xV_h}{V_0 - xV_{th}} \right] \right]^{-1}, \quad (7)$$

where $x = (R_L/R_{off} + 1)$, and for the circuit in Fig. 5b:

$$F_0 \sim \frac{I_0}{C(V_{th} - V_h)}. \quad (8)$$

Table 2 gives estimates of the maximum natural frequency $F_0^{10nF}$ when changing the switch scale, calculated from Equation (4), at $C = 10$ nF and $I_0 = I_{NDR}$. One can see that, as the switching structure size decreases, the oscillation frequency slightly falls. This fact seems contradictory, nevertheless it is accounted for by the drop in the value of the supply current $I_0 = I_{NDR}$, at which relaxation oscillations can still arise, and the capacitance is charged more slowly with decreasing $I_0$. On the other hand, the smaller the structure size is, the shorter the times $\tau_f$ and $\tau_{off}$ (see Table 1) are, which should lead to an increase in the switch operation frequency range. Indeed, if we assume that the minimum period of the oscillations is estimated as the sum of $\tau_f$ and $\tau_{off}$, $T_{min} = \tau_f + \tau_{off}$, then the maximum oscillation frequency can be estimated as $F_{max} = 1/T_{min}$, see Table 2. Thus, with a size $a$ decrease from 20 μm to 10 nm, one can expect an increase in $F_{max}$ by almost four orders of magnitude, up to about 300 MHz.

**Table 2.** Simulation data of the ultimate frequencies of the VO$_2$ based oscillator depending on the scaling parameter $a$ and film thickness $d$.

| $a$, μm | $d$, nm | $I_{NDR}$, A | $F_0^{10nF}$, kHz | $F_{max}$, MHz | $C_{min}$, F |
|---|---|---|---|---|---|
| 20 | 100 | 5.1·10⁻³ | 15 | 0.070 | 3.1·10⁻⁹ |
| 10 | 100 | 3.4·10⁻³ | 15 | 0.188 | 1.1·10⁻⁹ |
| 5 | 100 | 2.2·10⁻³ | 14 | 0.433 | 4.7·10⁻¹⁰ |
| 2 | 100 | 1.5·10⁻³ | 16 | 1.433 | 1.7·10⁻¹⁰ |
| 1 | 100 | 9.5·10⁻⁴ | 16 | 4.102 | 5.6·10⁻¹¹ |
| 0.5 | 100 | 4.4·10⁻⁴ | 11 | 7.625 | 2.1·10⁻¹¹ |
| 0.1 | 100 | 8.4·10⁻⁵ | 5.2 | 18.91 | 4.2·10⁻¹² |
| 0.01 | 20 | $1.1 \cdot 10^{-5}$ | 1.1 | 308.6 | $5.5 \cdot 10^{-14}$ |
| 1 | 20 | $5.1 \cdot 10^{-4}$ | 3.2 | 5.563 | $8.6 \cdot 10^{-12}$ |
| 10 | 20 | $1.6 \cdot 10^{-3}$ | 1.3 | 0.174 | $2.6 \cdot 10^{-10}$ |

Direct calculation of several periods of self-oscillations in the circuit shown in Figure 2 gives similar values of $F_{max}$. The simulation difficulty rises because of the need to select the minimum value of the capacitance $C_{min}$ at which the circuit is still able to oscillate, and this takes a long calculation runtime. Thus, we have proposed a simpler method for estimating $F_{max}$ using the switching-on $\tau_f$ and $\tau_{off}$ times extracted from a quasistatic I-V curve, either experimental or model one. Using known values of $F_{max}$, one can estimate $C_{min}$ from formula (4), which is presented in Table 2.

Table 2 shows that $F_{max}$ depends only slightly on the film thickness $d$, which is apparently related to the crucial role of heat dissipation into the substrate; that is why the film thickness has little effect on the channel heating and cooling times.

### 3.3. System of two C-coupled oscillators at weak coupling

Figure 7 shows the experimental circuit of two oscillators connected by the capacitive coupling.

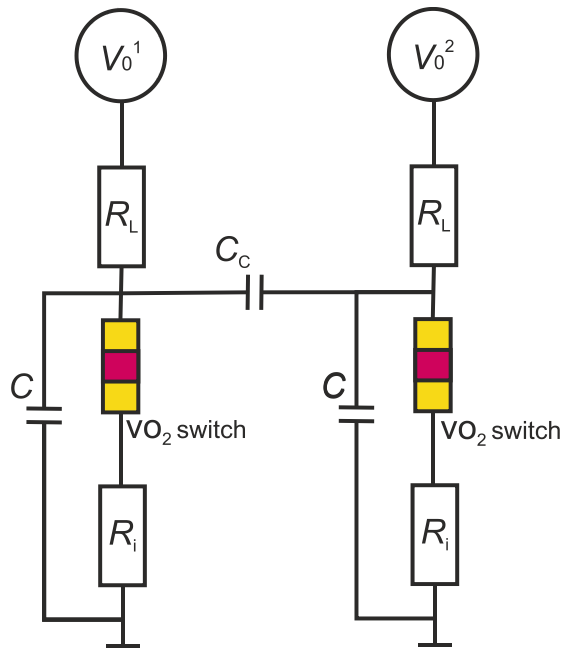

**Fig. 7**. Equivalent circuit of two oscillators with capacitive coupling: supply voltages – $V_0^1 = 82$ V, $V_0^2 = 82$ V; load resistors – $R_L = 50$ kΩ; parallel capacitances – $C = 100$ nF; current resistors – $R_i = 10$ Ω; $C_C$ – variable coupling capacitance.

By the term "weak coupling" we mean the condition $C_C < C$, while the switches can be located on the same substrate, but at a sufficiently far distance from each other, exceeding the action radius of the thermal coupling $R_{TC}$ described in Section 3.4 below.

At $C_C < 2.5$ nF, the oscillators behave as uncoupled ones. Oscillation waveforms and spectra of each oscillator qualitatively coincide with those of a single oscillator (Fig. 6) with the first harmonic frequencies $F^{1(1)} \sim 6.6$ kHz and $F^{2(1)} \sim 6.9$ kHz. At $C_C \geq 2.5$ nF, synchronization occurs. Figure 8b shows the current waveforms of oscillation and phase portraits of coupled oscillators with $C_C = 4$ nF. The spectra and phase portraits (Fig. 8 a,b) demonstrate synchronization in phase and frequency ($F^{1(1)} = F^{2(1)} = 6$ kHz) with a stable phase trajectory. The phase difference is $\Delta\varphi \sim 30^0$.

With increasing $C_C$, the natural frequencies ($F^{1(1)}$, $F^{2(1)}$) of oscillators are reduced, while the phase and frequency synchronization remains and the $\Delta\varphi$ value increases. As an example, Figure 8c shows the results of measurements for $C_C = 48$ nF. The oscillators are synchronized in the state close to antiphase ($\Delta\varphi > 160^0$). For example, for $C_C = 48$ nF, the first harmonic frequencies are $F^{1(1)} = F^{2(1)} = 4.6$ kHz, and the spectrum shape is similar to that at $C_C = 4$ nF (Fig. 8a).

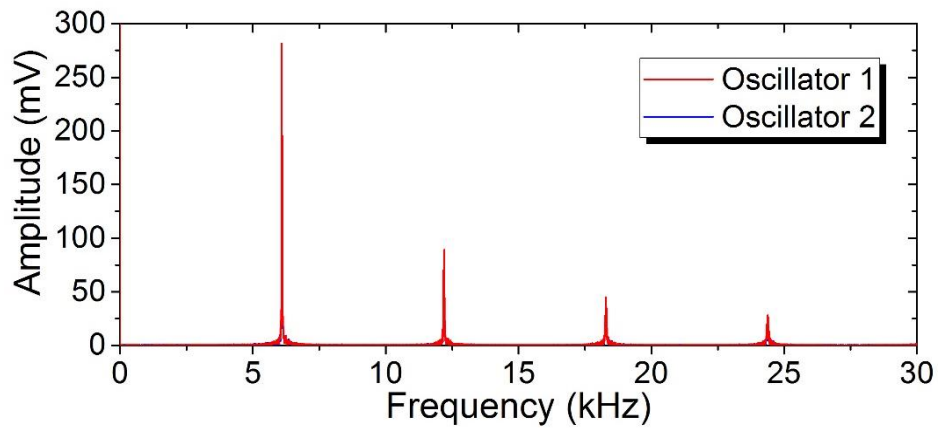

(a)

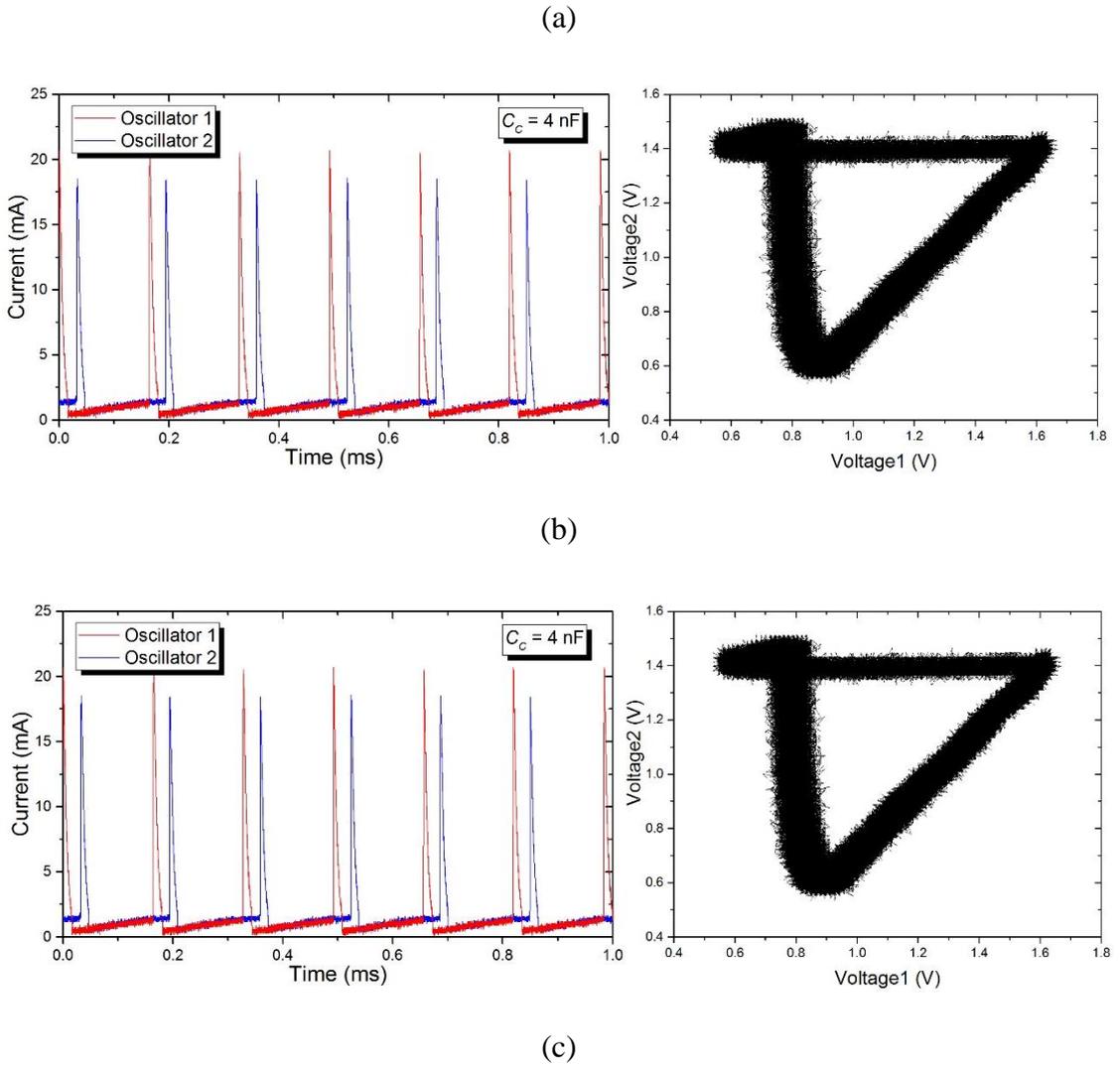

(b)

(c)

**Fig. 8**. (a) Experimental voltage spectra and current waveforms of C-coupled oscillators and their phase portraits for $C_C$ = 4 nF (b) and $C_C$ = 48 nF (c).

Figure 9 shows the dependences of the oscillator circuit first harmonic frequencies ($F^{1(1)}$, $F^{2(1)}$) and the phase difference $\Delta\varphi$ on the value of the coupling capacitance. The minimum value of the coupling capacitance, at which the synchronization occurs, is $C_C \sim 2.5$ nF. Within the 10 to 40 nF $C_C$ range, a linear dependence of the frequency on the coupling capacitance is observed, which is a feature of coupled oscillators, in contrast to single oscillators, where the frequency varies as $1/C$, see Equations (7) and (8). The influence of the coupling capacitance on the frequency arises because of the effective increase in the parallel capacitance $C$ in the circuit.

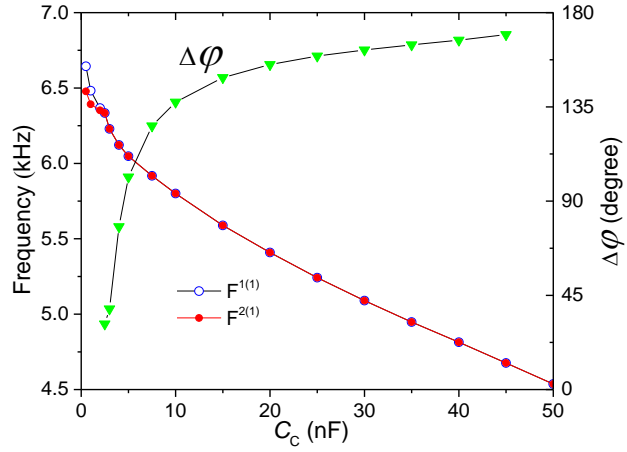

**Fig. 9**. Dependence of frequency and phase difference $\Delta\varphi$ of the oscillator circuits on the value of coupling capacitance $C_C$.

In the weak coupling regime, as the coupling capacitance increases, the phase difference $\Delta\varphi$ increases from 30 to 170º. Therefore, such a type of coupling might be used to design an ONN with phase keying.

At scaling the switch's size down, as shown above, the oscillation frequency of several tens or even hundreds of megahertz can be achieved. In this case, the synchronization phenomenon and the influence of the $C_C$ value on $\Delta\varphi$ take place, and the only factor that hinders the ONN miniaturization is the use of rather bulky elements, such as capacitors. Nevertheless, to obtain higher frequencies a lower capacitance $C$ is required, and the capacitances of the connecting wires can serve as such elements.

Another positive aspect of the frequency increase is the decrease in the time to enter the synchronization mode $T_S$. One can define and calculate this time as the time to synchronize at the simultaneous switching-on of the sources $V_0^1$ and $V_0^2$. Simulation of the circuit of Figure 7 in the LTspice software, using the threshold parameters listed in Table 1, shows how $T_S$ depends on the scaling parameter $a$, provided that $C = C_{min}$ and $C_c = C_{min}/10$. The results are presented in Figure 10.

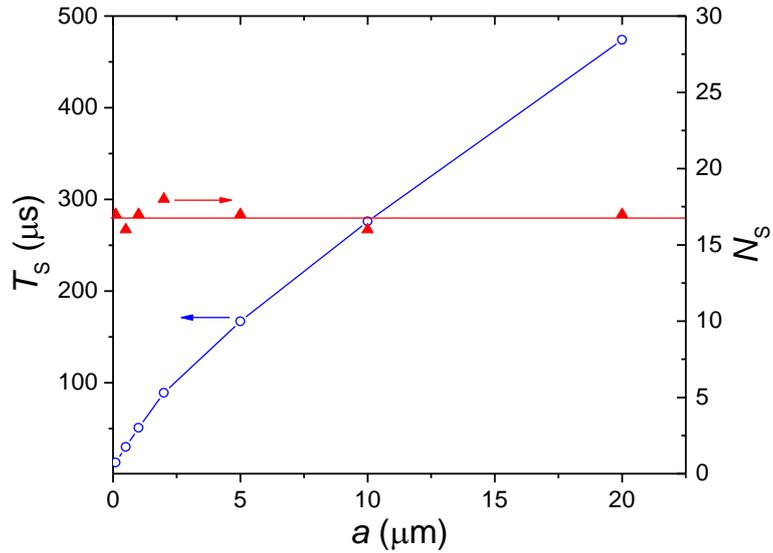

**Fig. 10**. Dependences of the synchronization time $T_s$ and the number of periods $N_s$ of synchronization on the channel scaling parameter $a$ ($C = C_{min}$, $C_c = C_{min}/10$).

The simulation results indicate a substantial decrease in $T_s$ with a decreasing of the switch dimensions, which is a positive factor for the ONN operation. During the $T_s$ calculation process, an interesting fact has also been revealed that the number of oscillation periods $N_s$ necessary for synchronization is practically independent of the $a$ value (see Figure 10).

### 3.4. System of two T-coupled oscillators at weak coupling

The thermal coupling (T-coupling) has first been described in our work [18]. In contrast to the C-coupling, the influence of one oscillator on another in this case is due to the mutual exchange of heat flows. This type of coupling is more promising for the implementation of the "all with all" coupling topology and it does not require the use of capacitors, thereby making this type of coupling more convenient for the ONN elements' scaling. The term "weak coupling" is used here when the distance between the switches on the substrate is greater than the characteristic dimensions of the switching channel $l$ and $h$.

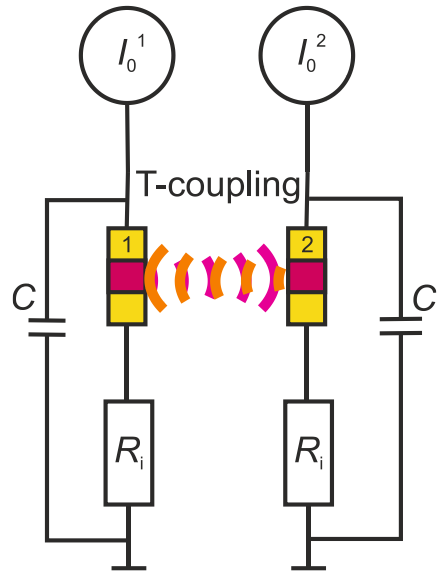

**Fig. 11**. Equivalent circuit of two oscillators with thermal coupling: current supply $I_0$, parallel capacitances $C = 100$ nF, and current resistors $R_i = 10$ Ω.

The oscillation observation circuit is shown in Fig. 11. The switches are located on a sapphire substrate, at a distance of 21 μm, as shown in Fig. 1a, yet they are galvanically isolated and the only coupling element here is the T-coupling. The current-voltage characteristics of the switches are similar to those shown in Fig. 3 (experimental curve).

External capacitances $C$ are connected to the switches in parallel and the power of the circuits is provided by current sources $I_0$. If the operating point of the circuit lies within the NDR region ($I_{th} < I_0 < I_h$), the switch voltage becomes unstable and self-oscillation occurs. The dependence of the single oscillator's oscillation frequency $F_0$ on the supply current $I_0$ is practically linear, in agreement with Equation (4).

To study the dynamics of a system of two thermally coupled planar oscillators, the current of one of oscillators $I_0^1$, as a variable parameter, is changed in the range of 660-820 μA, while the current of the second oscillator remains constant $I_0^2 = 720$ μA. Figure 12 shows a set of paired spectra of oscillations with a step of $\Delta I_0^1 = 10$ μA.

Initially, at $I_0^1 = 660$ μA the first harmonic frequencies $F^{1(1)}$ and $F^{2(1)}$ differ considerably, the oscillators operate independently and their frequencies are close to the natural ones. The spectra resemble the single oscillator spectra, possessing a broadening at the fundamental harmonic and a weak peak at the frequency of the nearby oscillator. This peak is due to the contribution of the thermal coupling at which the sporadic mutually induced switching events occur.

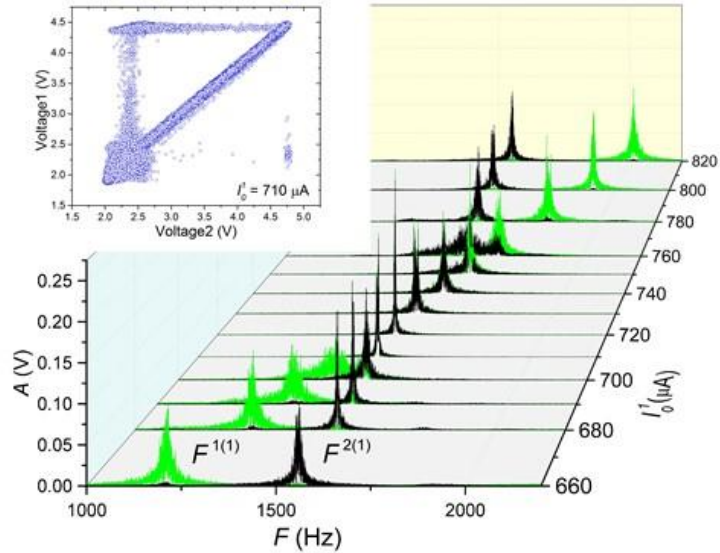

**Fig. 12.** Set of paired spectra of "variable" $F^{1(1)}$ (green graphs) and "stationary" $F^{2(1)}$ (black graphs) oscillators at varied supply current $I_0^1$ of "variable" oscillator and $d = 21$ μm.

Within the current range 710 μA $\leq I_0^1 \leq$ 750 μA, oscillators synchronize in frequency ($F^{1(1)} = F^{2(1)}$), and this effect is observed in the frequency range of 1550 to 1650 Hz. At $I_0^1 = 710$ μA, the best synchronization is observed, the effect of narrowing of the first harmonic spectrum width for both oscillators is noticeable, which is due to the ordering of the switching moments. Frequency synchronization is accompanied by the phase locking Δφ=0, and on the phase portrait (see insert in Fig.12), the attractor has the shape, predominantly, of an inclined straight line.

At synchronization of the fundamental frequencies, we often observe Δφ≈0, which is a distinctive feature of the thermal coupling caused by the switching initiation effect due to the induced temperature from the adjacent switch.

At $I_0^1 \geq 760$ μA, the system exits in the synchronization mode, and as the $I_0^1$ increases, the oscillators begin to behave as uncoupled. Thus, in the weak-coupling regime, the mutual influence of the oscillators affects their oscillation spectra only if the main harmonic frequencies approach each other.

In general, the mechanism of thermal coupling results from mutual heating of the areas in the neighboring switching structures. When the parallel capacitance $C$ is discharged, there should be a peak of the current leading to a local heating of the oscillator switch channel, which in turn causes a change in the neighbor switch temperature $\Delta T$. The dependence $\Delta T$ depends on many parameters, particularly, on the capacitance $C$, current resistor $R_i$, distance between switches, and the substrate thermal conductivity and heat capacity. Some of these parameters ($C$, $R_i$) can somehow be changed during the oscillator operation, while others are set at the structure fabrication stage. In our previous work [18], it has been shown that the effective oscillators' interaction radius $R_{TC}$ varies with variations in the values of capacitance $C$ and current resistance $R_i$ in the circuit. Here, it is to be clarified that by $R_{TC}$ we mean an average radius at which the peak temperature change, caused by electrical switching when the capacitor is discharged, reaches a quite determined value $\Delta T > 0.2$ K [18]. The value of $R_{TC}$ falls with increasing $R_i$ and with decreasing $C$. This is due to a decrease in the energy released on the switches at the time of discharge of the capacitor. When changing the switching channel dimension, one should also expect a decrease in the $R_{TC}$ value.

To present the thermal coupling mechanism in more detail Fig. 13a shows the results of numerical simulation of temperature distribution in the plane of two switches for the system when only one of oscillators operates 'VO$_2$ Switch 1' (switch parameters correspond to the experimental conditions, see Fig. 1a). Fig. 13a demonstrates the boundary where the temperature change $\Delta T = 0.2$ K restricts the area of thermal coupling of a single oscillator and determines its effective radius $R_{TC}$~32μm. The temperature in the center of the switch (point $x_1$) changes according to the time dependence shown in Fig. 13c (curve 1). The maximum amplitude of temperature impulses reaches high values $\Delta T(x_1)$ ~122 K. The boundary corresponding to $\Delta T = 0.7$ K passes through the center of the inactive switch 'VO$_2$ Switch 2' (point $x_2$), where the time diagram of temperature corresponds to Fig. 13c, curve 2. Therefore, induced from 'VO$_2$ Switch 1' temperature pulsation $\Delta T(x_2) = 0.7$ K in the channel area 'VO$_2$ Switch 2' is responsible for the effect of mutual synchronization. It should be noted that higher temperature change $\Delta T(x_3) = 1.8$ K (curve 3) is observed at point $x_3$ located between the switches.

We showed in [18] that when oscillation occurs one of the oscillators is the leader, i.e. its switching initiates switching of the neighbor switch. Fig. 13b shows temperature distribution in the case of synchronous operation of two oscillators when 'VO$_2$ Switch 1' is the leading one that matches the operation mode at $I_0^1 = 710$ µA, see Fig.12. This figure shows the moment when 'VO$_2$ Switch 1' is almost completely engaged into operation but 'VO$_2$ Switch 2' is just starting to operate with a small ~3µs time delay. It can be seen in comparison with a single oscillator operation when the boundary where $\Delta T = 0.2$ K broadens and the effective radius of two oscillators increases $R_{TC}$~44 µm. Also the temperature timing diagram at point $x_3$ (Fig.13c, curve 3') shows that the pulsation amplitude increases to $\Delta T(x_3) =$5.3K, that is apparently conditioned by superposition of induced temperatures from two oscillators.

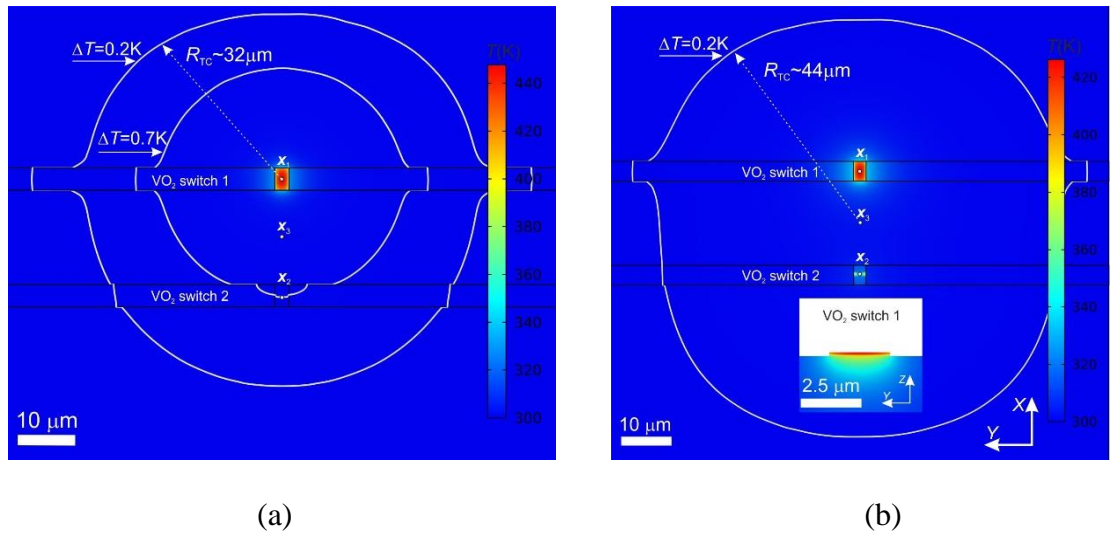

(a)          (b)

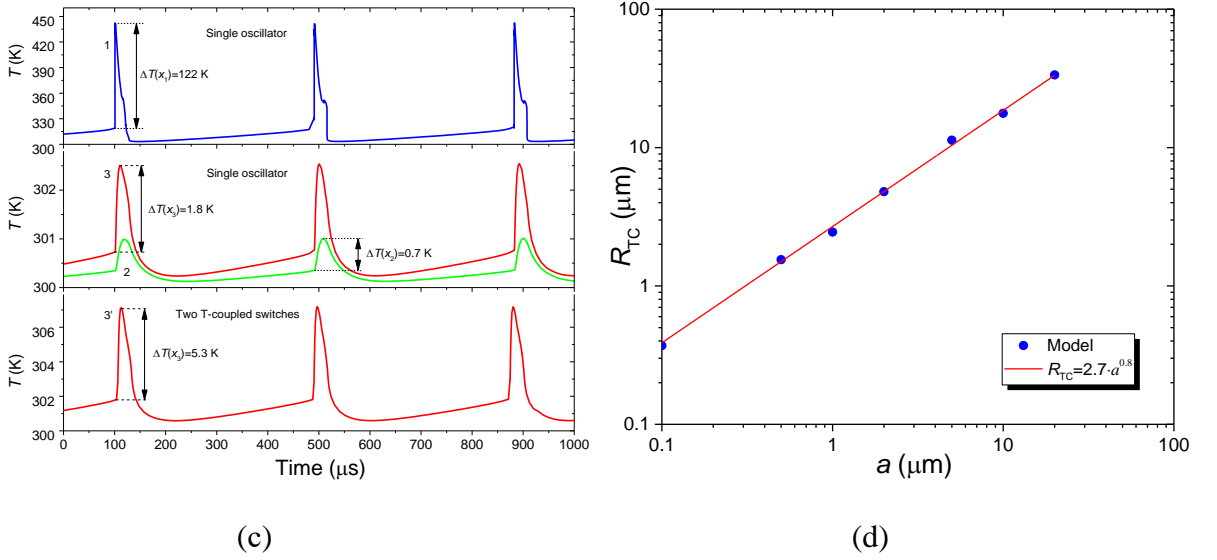

(c)                  (d)

**Fig. 13.** Results of numerical simulation of temperature distribution in the channels of a single (a) and two T-coupled switches (b); a) corresponding time-temperature dependences for a single oscillator at points $x_1$ (curve 1), $x_2$ (curve 2) and $x_3$ (curve 3) and for two T-coupled switches at point $x_3$ (curve 3'); d) The value of $R_{TC}$ as a function of scaling parameter $a$.

The thermal coupling action radius for a single oscillator as a function of the scaling parameter $a$ is shown in Fig. 13d; for the unification of the results, the calculations have been performed at frequencies $F_{max}$ for a given scale (see Table 2). It can be seen that $R_{TC}$ falls with decreasing $a$.

Alike the C-coupling case, the number of oscillation periods $N_s \sim 5$ necessary for synchronization is practically independent of the $a$ value. It is obvious that the time $T_s$ will depend on the magnitude of the thermal coupling, characterized, in our opinion, by the maximum induced temperature at the neighboring switch.

It should be remembered that the temperature change in the switching area is observed in the substrate volume (see. Fig 13.b, insert), therefore the radius of thermal coupling $R_{TC}$ restricts the low semi-sphere for the case in Fig. 13 a,b. Consequently, bulk property of thermal coupling provides some perspective for implementation of 3D integrated oscillatory neural network.

## 4. Conclusion

We have studied electrical switching with S-shaped *I-V* characteristics in two-terminal MOM devices based on vanadium dioxide thin films. By the *I-V* curve numerical simulation, it was shown that switching is associated with the thermally induced metal-insulator phase transition in vanadium dioxide. The behavior of the $VO_2$ switching devices, as well as the oscillators based on them and coupled oscillator circuits, was predicted when scaling the switch channel area dimensions.

The switching channel size varied in the range 10 nm to 20 μm. When calculating the temperature distribution in nanometer areas of 10-500 nm, we applied the classical heat equation (used in the Comsol platform) and electrical (thermal) constants for bulk samples and believed that this approach was adequate and yielded results corresponding to experiment. Corrections for the values of thermal and electric constants should be taken into account in the case if the characteristic structure size is comparable with the mean free path of an electron $l_e$ or phonon $l_{ph}$. For vanadium dioxide at room temperature, the estimates are $l_e \sim 0.3$ [19] nm and $l_f \sim 10$ nm [20], which are much smaller than the channels sizes used in calculations.

The simulation revealed a certain oscillator critical frequency $F_{max}$, which could reach the value of 300 MHz for the $VO_2$ channel sizes of 10×10 nm and film thickness ~20 nm. The minimum switching time was calculated to be about 3 ns in this case which is comparable with the data presented in the literature, ~9 ns [2]. Further decrease in the switching channel size seemed to be hardly feasible because of the MIT suppression due to the dimensional effects, as was discussed in Section 3.1.

Relaxation oscillations have been studied in circuits with $VO_2$-based switches. The dependences of the oscillator critical frequency $F_{max}$, threshold power and voltage, as well as the time of current rise, on the switching structure size are obtained by the numerical simulation. An empirical relation of the threshold voltage versus switching area size and film thickness is found, which allows for a preliminary estimation without a prolonged time-consuming calculation. In addition, this relation reveals the fact that the ratio $V_{th}/V_h$ = const ≈ 4.22 is independent of $d$ and $a$. Analytical approximation formulas (3)-(6) define the physics of empirically obtained coefficients and power dependences (1)-(2). Next, it is shown that oscillatory neural networks can be implemented on the basis of coupled $VO_2$ oscillators. For the weak C-coupling, we have found the dependence of the phase difference upon synchronization on the coupling capacitance value. When the switches are scaled down, the limiting time of synchronization is reduced to $T_s$ ~13 μs, while the number of oscillation remains unchanged $N_s$ ~17. In case of the weak thermal coupling in the synchronization mode, we observe in-phase behavior of oscillators, and there is a certain range of parameters of the supply current, within which the synchronization effect becomes possible. With a decrease in the scaling parameter (characteristic structure size), a decrease in the thermal coupling action radius is observed, which can vary in the range 0.5-50 μm for structures with characteristic dimensions of 0.1 - 5 μm, respectively.

Thermal coupling has some limitations. First, thermal coupling force depends on the mutual distance which is predetermined only at the manufacturing stage. Nevertheless, we showed in [18] that it is possible to control the value of $R_{TC}$ and therefore the coupling force by varying $R_i$ and $C$. Second, switches parameters depend on the changes of the environmental temperature $T_0$ and the switching effect degenerates if $T_0>T_{th}$. However, the latter limitation is common for all elements having S-shape *I-V* curve and it could be eliminated either by outside thermostatic control or by application of materials with higher $T_{th}$, such as $NbO_2$ [21].

Comparison of thermal and capacitance coupling demonstrates that the effect of thermal coupling is the basic one and it always occurs if the distance between the switches is less that $R_{TC}$. Therefore if our purpose is to realize only capacitance coupling we should arrange switches at the spacing that exceeds $R_{TC}$. However, in this case the packaging density of oscillators decreases markedly for ONN fabrication. Consequently, thermal coupling is more promising, allows for coupling capacitors exception and provides decoupling of direct and alternate currents. Moreover, the thermal coupling allows for oscillators spacing at the distance less than $R_{TC}$, this being the essential requirement. Bulk property of thermal coupling provides some perspective for implementation of 3D integrated oscillatory neural network. These advanced features increase the possibilities of ONN nano scaling and integration.

Thus, $VO_2$ is a vivid example of how strongly correlated materials might be efficiently used in oxide electronics, encompassing such its diverse areas as, for example, high-speed electronic switches and relaxation oscillators as elements of ONNs.

**Acknowledgments**

This work was supported by Russian Science Foundation, grant no. 16-19-00135.